
\ifx\mnmacrosloaded\undefined \input mn\fi
%

\newif\ifAMStwofonts
\ifCUPmtplainloaded \else
  \NewTextAlphabet{textbfit} {cmbxti10} {}
  \NewTextAlphabet{textbfss} {cmssbx10} {}
  \NewMathAlphabet{mathbfit} {cmbxti10} {} 
  \NewMathAlphabet{mathbfss} {cmssbx10} {} 
  \ifAMStwofonts
    \NewSymbolFont{upmath} {eurm10}
    \NewSymbolFont{AMSa} {msam10}
    \NewMathSymbol{\upi}     {0}{upmath}{19}
    \NewMathSymbol{\umu}     {0}{upmath}{16}
    \NewMathSymbol{\upartial}{0}{upmath}{40}
    \NewMathSymbol{\leqslant}{3}{AMSa}{36}
    \NewMathSymbol{\geqslant}{3}{AMSa}{3E}

     \let\le=\leqslant
     \let\ge=\geqslant
  \else
    \def\umu{\mu}
    \def\upi{\pi}
    \def\upartial{\partial}
  \fi
\fi


\pageoffset{-1.5pc}{+3pc}

\loadboldmathnames



\pagerange{000--000}    
\pubyear{1995}
\volume{000}

\def\CaII{Ca~{\sevenrm II}\ }

\begintopmatter  

\title{A sample-oriented catalogue of BL Lacertae objects}
\author{Paolo Padovani$^1$ and Paolo Giommi$^2$}
\affiliation{$^1$ Dipartimento di Fisica, II Universit\`a di Roma `Tor
Vergata', Via della  Ricerca Scientifica 1, I-00133 Roma, Italy}
\medskip
\affiliation{$^2$ ESIS, Information Systems Division of
ESA, ESRIN, Via G. Galilei, I-00044 Frascati, Italy}

\shortauthor{P. Padovani and P. Giommi}

\shorttitle{A catalogue of BL Lacs}


\acceptedline{Accepted 1995 July 27. Received 1995 July 27; in original form
1995 April 19}

\abstract {We present a catalogue of 233 BL Lacertae objects compiled through
an extensive bibliographic search updated to mid-1995. A large fraction of
the sources listed in the catalogue belongs to well-defined samples and can be
used for statistical purposes. A smaller fraction consists of miscellaneous
(but confirmed) BL Lacs and of objects classified as BL Lac candidates. We
discuss the selection criteria of the different samples, report the
discovery of two previously unnoticed BL Lacs in the Palomar--Green survey,
and comment on the possible association of some of the still unidentified high
galactic latitude gamma-ray (EGRET) sources with BL Lacs. Some statistical
properties of the catalogue are also briefly discussed.}


\maketitle  

\section{Introduction}

BL Lacertae objects are hard to find. This is simply due to one of their
defining features: the almost complete lack of emission lines.
As a result, contrary to most other astronomical sources, only a few objects
of this class have been discovered at optical frequencies.
About 95 per cent of known BL Lacertae objects have in fact
been discovered in the radio or X-ray band, where they can be more easily
recognized thanks to other distinguishing properties such as flat radio
spectra and a distinctive multifrequency energy distribution. BL Lacertae
objects are also intrinsically rare and constitute only a few per cent of the
known population of active galactic nuclei (AGN).

The need for an up-to-date BL Lac catalogue came for a practical reason. We
were studying the AGN content of the WGA catalogue (White, Giommi \& Angelini
1994), a large catalogue of X-ray sources generated from all the
{\it ROSAT}~PSPC
pointed observations. We wanted to extract from it all known BL Lacertae
objects to analyse their X-ray properties, but, when we started using the
V\'eron-Cetty \& V\'eron (1993a) and Hewitt \& Burbidge (1993) catalogues, we
realized that the number of BL Lacs had recently increased in
such a way as to require a new compilation. Also, the criteria according to
which
objects had been called BL Lacs in previous catalogues were highly
inhomogeneous.

We then put together a list of BL Lacertae objects, taking a novel approach.
Instead of assembling all objects ever called BL Lacs in the literature, we
started from the (by now quite numerous) complete samples, adding at the end
additional objects from existing catalogues and the literature. By {\it
complete sample}~here we mean an homogeneous set of sources detected in a {\it
statistically well-defined}~and {\it completely identified}~survey (although
in some cases the identification process is not yet 100 per cent complete). In
most cases statistically well-defined means flux limited in one or more energy
bands.

The result of this effort is a catalogue of 233 BL Lacs, i.e. significantly
larger than the V\'eron-Cetty \& V\'eron (171 BL Lacs) and Hewitt \& Burbidge
(90 BL Lacs) catalogues but above all, we believe, based on more homogeneous
criteria. Most samples, in fact, adopt a classification based on equivalent
width $W_{\lambda}$, 5 \AA~being the dividing line between BL Lacs and
quasars. Although there are undoubtedly borderline objects, in which emission
lines appear when the continuum is in a low state, this value seems to
separate quite well the two classes (see discussion in Urry \& Padovani
1995).

The structure of the paper is as follows: in Section 2 we describe the
catalogue
and in particular the samples on which it is mostly based, while in Section 3
we
present some of the statistical properties of the catalogue.

\section{The catalogue}

The catalogue includes all objects from the BL Lac samples known to us at
the time of writing (1995 June) plus objects listed in the V\'eron-Cetty
\& V\'eron (1993a) and Hewitt \& Burbidge (1993) catalogues as follows:
objects common to the two catalogues were included under the label
`miscellaneous'; objects belonging to only one of the two were
conservatively labelled `candidate'.
This last group includes also BL Lac candidates belonging to
complete samples or found in the literature. We have excluded
from our list objects originally classified as BL Lacs or BL Lac candidates
but which were later shown to have strong lines: examples are 2201+044,
included in the {\it HEAO}-1 sample by Laurent-Muehleisen et al. (1993) but
recently shown to be a Seyfert 1 galaxy (V\'eron-Cetty \& V\'eron 1993b),
and 1214+1753, a BL Lac candidate in Foltz et al. (1987) but a broad absorption
line QSO in Stocke et al. (1992).

The catalogue, which is `complete', to the best of our knowledge, as far as
BL Lacs in samples and miscellaneous objects are concerned, is presented in
Table 1. Column 1 gives the most common name(s), columns 2 and 3 the J2000
positions for each object. Columns 4, 5 and 6 give
the redshift, $V$ magnitude and radio flux at 5 GHz, while column 7 contains
the
references for these quantities. Finally, in column 8 we give references to
X-ray data while in column 9 we report the sample(s) to which the object
belongs or the `miscellaneous' or `candidate' classification (the latter
followed, if applicable, by the name of the sample to which the object
belongs). The data generally come `as given' by the paper describing the
sample or from the catalogues. When radio fluxes were not provided, we
searched available 5-GHz radio catalogues at northern and southern
declinations (Becker, White \& Edwards 1991; Wright et al. 1994; Griffith et
al. 1994, 1995), which at
present reach $\sim 20 - 40$ mJy. It can then be safely assumed that the 13 BL
Lacs in this catalogue lacking radio data have 5-GHz radio fluxes below these
limits. Positions are usually good to within a few arcseconds for the radio-
and optically-selected sources. A similar accuracy is reached also by the
X-ray-selected sources in the {\it Einstein}~Imaging Proportional Counter
(IPC) Slew survey, for most of which precise radio positions are given in
Perlman et al. (1995b), in the {\it Einstein} Observatory Extended Medium
Sensitivity Survey (EMSS), whose optical coordinates have been taken from
Maccacaro et al. (1994), and in the {\it ROSAT} all-sky survey (RASS), whose
BL Lacs
have optical coordinates (Bade, Fink \& Engels 1994). We also present here
previously unpublished radio positions, good to within 1 -- 2 arcsec, for the
following {\it EXOSAT} BL Lacs: EXO0706.1+5913, EXO0811.2+2949, EXO1004.0+3509,
EXO1118.0+4228, ~EXO1146.9+2455, and EX\-O1811.7+3143. In a few cases accurate
radio positions were obtained from the NASA/IPAC Extragalactic Database (NED).
In summary, coordinates are uncertain up to a few tens of arcseconds only for
four {\it EXOSAT} BL Lacs (EXO0044.4+2001, EXO0423.4$-$0840, EXO0556.4$-$3838,
and
EXO1415.6+2557), one {\it HEAO}-1 BL Lac (1H 0829+090, plus 1H 1914$-$194, for
which coordinates are not available), and the five new RASS BL Lacs published
by Brinkmann et al. (1995). In cases when an object belongs to more than one
sample, the most accurate coordinates are quoted.

One of the characteristics of BL Lacs is their strong variability, which
generally increases with frequency. While this is not a problem in the radio
band where large-amplitude variability is rare (e.g. Miller \& Wiita 1991),
$V$ magnitudes should be taken with care. As regards the 1-Jy BL Lacs, however,
we provide values which should be quite representative of the `typical'
state of the objects, from Padovani (1992). The reader is referred to that
paper for a description of the derivation of these magnitudes and for the
appropriate references. We also note that optical monitoring data for many BL
Lacs are available in the literature (e.g. Pica et al. 1988; Webb et al.
1988; Falomo, Scarpa \& Bersanelli 1994).

Given that a number of additional complications are present in
the X-ray data (different instruments and bands, uncertainties in the spectral
indices, photoelectric absorption), we do not provide
X-ray fluxes. We prefer instead to give at least one reference, so that
X-ray data can be easily retrieved. X-ray references have been compiled as
follows: our main sources were the catalogue of Della Ceca et al. (1990), a
compilation of X-ray data up to 1986, and the catalogue of X-ray spectra of
Ciliegi, Bassani \& Caroli (1993), complete up to the end of 1991. We also
refer to the
recent WGA catalogue of {\it ROSAT} PSPC sources (White et al. 1994),
while in the case of X-ray-selected objects we refer to the original papers
describing the sample. Finally, for those sources not included in the
above-mentioned catalogues, we first performed a literature search and then
searched
the {\it Einstein} IPC and {\it EXOSAT} databases. At the end of this process,
30
objects (or about 13 per cent of the catalogue) had no X-ray references.
Considering only confirmed BL Lacs this number goes down to 10 (or about 5 per
cent of the confirmed objects).

The catalogue breaks up into the following sub-classes: confirmed BL Lacs in
samples (159 objects or 68 per cent of the catalogue), miscellaneous BL Lacs
(24
objects or 10 per cent of the catalogue), and BL Lac candidates or objects
whose BL
Lac classification is still uncertain (50 objects or 22 per cent of the
catalogue), for a total of 233 objects.

We now describe the various samples and classes, grouping them into wavelength
bands.

\subsection{Radio surveys and catalogues}

The radio band is where the class of BL Lacertae objects was discovered. For
many years the great majority of the BL Lacs known were found among the
sources detected in large radio surveys. As a consequence, all the classical
and well-studied objects have been discovered at these frequencies.

\subsubsection{The 1-Jy sample}

This is currently the largest complete radio sample of BL Lacs. Described in
Stickel et al. (1991) and Stickel, Fried \& K\"uhr (1993), it includes 34
objects extracted from the 1-Jy catalogue (radio flux $f_{\rm r} \ge 1$ Jy at
5 GHz:
K\"uhr et al. 1981a), a flux-limited catalogue which covers essentially the
whole sky excluding the galactic plane ($|b| < 10^{\circ}$) and the Magellanic
Clouds, according to the following criteria: (1) flat radio spectrum between
2.7 and 5 GHz ($\alpha \le 0.5$, $f_{\nu} \propto \nu^{-\alpha}$); (2)
magnitude brighter than 20; (3) emission lines in the optical spectrum absent
or weak with a rest-frame equivalent width of the strongest line $< 5$ \AA.
Three more objects are included in the updated version of the 1-Jy catalogue
(Stickel, Meisenheimer \& K\"uhr 1994). Two of these have $V \ge 20$, while
the third one (PKS 2149+173) has $V = 18.9$, although during the period of the
spectroscopic observations it was below the 20th magnitude limit (Stickel \&
K\"uhr 1993a). PKS 0521$-$365, an object classified as a BL Lac in the
literature but originally excluded from the 1-Jy sample because some of its
lines had $W_{\lambda} > 5$ \AA, has been included as an uncertain BL Lac,
since its [O{\sevenrm~III}]~luminosity is more typical of BL Lacs than of
quasars (see discussion in Urry \& Padovani 1995).

Brunner et al. (1994) have reported on a {\it ROSAT} observation of S5 0454+844
(which belongs both to the 1-Jy and S5 samples), which shows the BL Lac object
to be only 48 arcsec away from a source about five times brighter. They
therefore
suggest that previous reports of X-ray emission from this object (which are
referred to in Table 1) could be due to a misidentification.

It has been suggested (Perlman et al. 1995a) that the requirement on the radio
spectral index ($\alpha_{\rm r} \le 0.5$), imposed as a criterion for the
selection of the 1-Jy BL Lacs to exclude the bulk of the radio galaxies, might
have resulted in the loss of some objects, since BL Lacs with steeper radio
spectral indices are known. To estimate the magnitude of this
effect, we performed the following simple calculation: out of the 119
confirmed BL Lacs which, to our knowledge, have 2.7 -- 5 GHz spectral index
information, only 8 have $\alpha_{\rm r} > 0.5$. Out of these, one (S5
1749+701) was included nevertheless in the 1-Jy sample because the steep radio
spectrum was clearly due to variability and non-simultaneous measurements
(Stickel et al. 1991). As regards the remaining ones, variability cannot be
the explanation for the steepness of $\alpha_{\rm r}$ in the case of PKS
0548$-$322, MS1207.9+3945 and MS1402.3+0416 since the multifrequency radio
measurements were contemporaneous (Stocke et al. 1985), while it cannot be
excluded in the remaining four objects, ON 231, MS1407.9+5954, RXJ00079+4711
and RXJ16442+4546. We then estimate that, out of the 85 BL Lacs not included in
the 1-Jy, S4 and S5 samples (which were selected to have flat radio spectra),
between 3 (4 per cent) and 7 (8 per cent) have $2.7-5$ GHz spectral indices
steeper than 0.5. This suggests that the condition requiring a flat radio
spectrum might cause the loss of only 1 to 3 BL Lacs in the 1-Jy sample.

\subsubsection{The S4 sample}

The S4 sample includes 14 objects extracted from the S4 catalogue ($f_{\rm
r} \ge 0.5$ Jy at 5 GHz, $35^{\circ} \le \delta \le 70^{\circ}$ and $|b| \ge
10^{\circ}$: Pauliny-Toth et al. 1978; Stickel \& K\"uhr 1994) by Stickel \&
K\"uhr (1994) using the same criteria as those applied to 1-Jy BL Lacs. Note
that S4 1652+398 (Mrk 501) and S4 1823+568 have been mistakenly classified as
a normal galaxy and a QSO respectively in Stickel \& K\"uhr (1994): both are
in fact confirmed 1-Jy BL Lacs. About 10 per cent of the S4 sources are still
classified as empty fields so a small number of BL Lacs could still be
unidentified.

\subsubsection{The S5 sample}

The S5 sample includes 13 objects extracted from the S5 catalogue ($f_{\rm
r} \ge 0.25$ Jy at 5 GHz, $\delta \ge 70^{\circ}$ and $|b| \ge 10^{\circ}$:
K\"uhr et al. 1981b) by K\"uhr \& Schmidt (1990). The selection criteria are
slightly different from those adopted for the 1-Jy and S4 samples: they
include in fact maximum optical polarization $P_{\rm max}$ larger than 3 per
cent on at least one occasion, while it is not clear what is the equivalent
width limit adopted to separate BL Lacs from quasars. (Note that all but two
1-Jy BL Lacs have $P_{\rm max} > 3$ per cent [Stickel et al. 1994], although
this was not one of the selection criteria.) We have excluded S5 1053+81 from
the sample because its spectrum shows emission lines (Xu et al. 1994), and we
have added a BL Lac candidate, S5 2353+81 (Stickel \& K\"uhr 1993b). The S5
catalogue is currently being updated by Stickel \& K\"uhr (in preparation).
\par We note that the dynamical range of source flux in radio surveys is small
and of order $\approx 10$. This demonstrates that the study of BL Lacs is still
at a very early stage
even in the part of the electromagnetic spectrum where
these objects were first discovered more than 25 years ago.

\subsection{Optical surveys and catalogues}

Several surveys that make use of various detection methods have been
carried out in the optical band. Some of these have been tuned to the search
for BL Lacs. However, despite the large efforts only a handful of BL Lacs have
so far been discovered at these frequencies.

\subsubsection{The Palomar--Green sample}

The Palomar--Green (PG) sample covers 10714 deg$^2$ of sky down to an
average limiting magnitude $B = 16.1$ for $U-B < -0.46$ (Green, Schmidt \&
Liebert 1986). Four BL Lacs were initially identified in the catalogue: OJ
287, OQ 530, PG 1553+113, 1H 1219+301 (2A 1219+305), the last object not
belonging to the complete sample. Three more objects, originally misclassified
as white dwarfs, were discovered by Fleming et al. (1993) by cross-correlating
the PG white dwarf list with the RASS. To check if
there were other known BL Lacs still lurking in the catalogue we
cross-correlated the PG sample with the present BL Lac catalogue. Much to our
surprise, we found that Mrk 421 was only about 51 arcsec away from PG 1101+385
and PKS 2254+074 was about 9 arcsec away from PG 2254+074, the former being
classified as a `composite spectrum object' in Green et al. (1986), the
latter, which does not belong to the complete sample, being unclassified. While
an offset of 9 arcsec is consistent with the typical accuracy of the PG
positions (about 8 arcsec in each coordinate), the larger deviation of 51
arcsec could be due
to the fact that Mrk 421 is an extended object. Indeed, an examination of the
charts in the PG catalogue shows that PG 1101+385 coincides with Mrk 421 (R.
Green, private communication). The total number of BL Lacs in the PG sample is
then 9 objects, of which 7 belong to the complete sample. We cannot exclude the
possibility that other BL Lacs are present in the sample awaiting discovery.
The
PG sample gives only a lower limit to the number of optically selected BL
Lacs, since many such objects have $U-B$ colours above the selection limit:
using the colours tabulated in V\'eron-Cetty \& V\'eron (1993a) and Hewitt \&
Burbidge (1993), to derive the $U-B$ distribution of known objects, we
estimate that the PG sample misses about 40 per cent of BL Lacs.

\subsubsection{The optical variability sample}

A survey of optically variable quasars over 18 deg$^2$ was carried out by
Hawkins et al. (1991). This led to the discovery of two BL Lacs brighter than
$B \sim 19$; some more could be present in the field. $V$ magnitudes for the
two
objects have been derived from their mean $B$ magnitudes and $B-V$ colours.

\subsubsection{The optical polarization survey}

An optical polarization survey covering 560 deg$^2$ of high
galactic latitude sky was carried out by Jannuzi, Green \& French (1993).
Only one BL Lac
candidate was found and the conclusion of this work was that most BL Lacs
are not highly polarized (e.g. they do not spend much time at $P_{\rm
max} \ga 30$ per cent) and that much higher sensitivities in polarization
levels are needed to detect a significant number of new BL Lacs.

\subsection{X-ray surveys and catalogues}

Over the past few years several X-ray surveys have become available and
many new BL Lacs have been discovered. Thanks to their strong X-ray emission,
selection in this band has become the most effective way of discovering new BL
Lacs. At present more than 50 per cent of the BL Lacs known have been
discovered at X-ray frequencies.

\subsubsection{The EMSS sample}

The EMSS (Gioia et al. 1990; Stocke et al. 1991; Maccacaro et al. 1994) is a
flux-limited sample of X-ray sources discovered serendipitously in 1435 {\it
Einstein} IPC fields centred on high-latitude ($|b| > 20^{\circ}$) targets.
It covers 780 deg$^{2}$ in the 0.3 -- 3.5 keV band and goes down to $f_{\rm x}
\sim 5 \times 10^{-14}$ erg cm$^{-2}$ s$^{-1}$, albeit in a much smaller area
since the area of sky covered is a strong function of X-ray flux (see table 5
of Gioia et al. 1990).

The EMSS includes 34 BL Lacertae objects [plus two BL Lac candidates, which is
one less than in the original list, since
MS1332.6$-$2935, a BL Lac candidate in Stocke et al. (1991), is now a confirmed
BL Lac (Perlman et al. 1995b)], selected according to the following criteria:
{\it observed} equivalent width of any emission line $< 5$ \AA~and evidence
for dilution of starlight in the spectrum by a non-thermal continuum, which in
practice means a \CaII break with a relative flux depression blueward across
the break $\le 25$ per cent in the spectra (normal ellipticals would have
values
around 50 per cent). Note that, although the equivalent width division is the
same as the one applied to the 1-Jy and S4 BL Lacs, here it refers to the {\it
observed} and not to the {\it rest-frame} value. It then follows that, since
$W_{\lambda,\rm rest} = W_{\lambda,\rm obs}/(1+z)$, the EMSS could classify as
AGN some objects that would have been considered as BL Lacs by the 1-Jy
sample, especially at high redshifts.

Browne \& March\~a (1993) have suggested that the EMSS might misclassify some
low-luminosity BL Lacs whose light is swamped by the host galaxy, which is
typically a bright elliptical. Padovani \& Giommi (1995a), within their
hypothesis on the relationship between X-ray- and radio-selected BL Lac
objects, have performed numerical simulations to establish the incompleteness
level of the EMSS implied by the Browne \& March\~a effect. By applying the
prescription of Browne \& March\~a to establish if a BL Lac is recognized as
such or if it is misclassified,
they found that about 10 per cent of the EMSS BL Lacs could be lost. Perlman
et al. (1995a) have also discussed this effect and looked for possible
misidentifications of BL Lacs with clusters of galaxies, coming up with one
possible BL Lac (MS1019.0+5139) and four other (unlikely in their view)
possibilities. We note that PKS 2316$-$423 (MS2316.3$-$4222), classified as a
cluster of galaxies by Stocke et al. (1991), has recently been suggested to
host a BL Lac object by Crawford \& Fabian (1994).

Complete BL Lac subsamples from the EMSS have been presented by Morris et al.
(1991) and Wolter et al. (1994).

\subsubsection{The EXOSAT sample}

The {\it EXOSAT} High Galactic Latitude Survey (HGLS: Giommi et al. 1991) is a
flux-limited sample of X-ray sources discovered serendipitously in 443 Channel
Multiplier Array (CMA) fields centred on high-latitude ($|b| > 20^{\circ}$)
targets. It covers 783 deg$^{2}$ in the 0.05 -- 2.0 keV band and goes down to
$f_{\rm x} \sim 2 \times 10^{-13}$ erg cm$^{-2}$ s$^{-1}$, albeit in a much
smaller area since the area of sky covered is a strong function of X-ray flux
(see fig. 1 of Giommi et al. 1991).

The {\it EXOSAT} HGLS includes 12 BL Lacertae objects (plus two BL Lac
candidates)
selected according to criteria similar to those of Stocke et al. (1991).

\subsubsection{The {\it HEAO}-1 sample}

The {\it HEAO}-1 Large Area Sky Survey (LASS, also known as {\it HEAO} A-1)
has produced a catalogue of bright, hard X-ray (0.8 -- 20 keV) sources over
the entire sky (Wood et al. 1984). The optical identification programme is
still
on-going and the final BL Lac sample has not been published yet. Schwartz et
al. (1989) have presented some preliminary results, while Laurent-Muehleisen
et al. (1993) have published a list of 29 {\it HEAO}-1 BL Lacs, one of which
(2201+044) is now known to be a Seyfert 1 galaxy (V\'eron-Cetty \& V\'eron
1993b). PKS 0521$-$365, previously discussed, has been included in our list as
an uncertain BL Lac.

Although it is not entirely clear what were the precise criteria for the
classification as BL Lacs, it is known that the {\it HEAO}-1 BL Lacs have been
selected on the basis of their UV excess (Schwartz et al. 1989). This means
that the {\it HEAO}-1 sample suffers from the same incompleteness as
the PG sample, discussed above (Section 2.2.1).

Laurent-Muehleisen et al. (1993) give positions only for a subsample of their
objects. This was not a problem for the majority of the remaining sources,
since they are in common with other samples (Slew, 1-Jy). In one case (1H
0829+089) positions were derived from a search in radio catalogues around
the IAU position (i.e. the position obtained from the source name) and
therefore the coordinates are only good to within a few tens of arcseconds. In
another case (1H 1914$-$194) no radio source was found within a degree from
the IAU position, so no coordinates are available for this object.

\subsubsection{The Slew survey sample}

The IPC Slew survey has been constructed using the {\it Einstein} `slew'
data taken when the satellite was moving from one target to the next (Elvis et
al. 1992), and covers a large fraction of the sky with sensitivities $\sim 5
\times 10^{-12}$ erg cm$^{-2}$ s$^{-1}$, reaching a flux limit $f_{\rm x}
\la 10^{-12}$ erg cm$^{-2}$ s$^{-1}$ over a much smaller fraction of the
sky (Schachter, Elvis \& Szentgyorgyi 1993b).

Perlman et al. (1995b) (see also Schachter et al. 1993a) have presented a
sample of 62 BL Lacs (which include 2 probable BL Lacs) extracted from the
Slew survey adopting the same classification criteria as the EMSS survey. This
is the largest BL Lac sample so far. Out of these objects, a complete sample
of 48 BL Lacs has been defined. Six more objects have not been observed
spectroscopically but have broad-band energy indices typical of BL Lacs and
have been included here as BL Lac candidates. To these we also add
1ES1249+174W, discussed by Perlman et al. (1995b).

The {\it HEAO}-1 and the Slew surveys include many BL Lacs previously
selected in radio surveys. This shows a severe limitation of a classification
that has been frequently used in the recent past, based solely on the selection
band. This classification divides the class of BL Lacs into
{\it X-ray-selected} (or XBL) and {\it Radio-selected} (or RBL) depending on
the band
where the object was discovered: a number of objects could therefore be
classified as both XBL and RBL. To overcome this difficulty Padovani \&
Giommi (1995a) have proposed to classify the objects on the basis of the ratio
between the X-ray and radio fluxes (a parameter which univocally identifies an
object) and suggested a division into `RBL-like' or low-energy cutoff
BL Lacs (LBL) and `XBL-like' or high-energy cutoff BL Lacs (HBL). This
division corresponds in fact to a break in their broad-band spectrum at
infrared/optical frequencies for the former objects and at ultraviolet/X-ray
energies for the latter.

\subsubsection{The RASS sample}

The {\it ROSAT} all-sky survey (RASS; Voges 1992) includes about 60000 X-ray
sources, several hundred of which should be BL Lacs. The identification
programme will inevitably take several years but early results have started to
appear in the literature. Bade et al. (1994) report the discovery of
10 new BL Lacs through follow-up spectroscopy of AGN candidates detected in
the RASS survey. Three of the objects are in common with the Slew survey. All
optical spectra satisfy the EMSS BL Lac criteria of Stocke et al. (1991).
Brinkmann et al. (1995), quoting a private communication from A. Kock, present
five more new RASS BL Lacs which are (mostly) included in the 5-GHz survey of
Condon, Broderick \& Seielstad et al. (1989). Since no coordinates were given,
we obtained them from
the radio catalogue, searching near the coordinates obtained from the source
name: they should therefore be accurate only to within a few tens of
arcseconds. In one case (RXJ16264+3513), the radio source does not belong to
the 5-GHz catalogue and therefore no coordinates are available.

\subsubsection{The WGA and ROSATSRC catalogues}

Two catalogues of {\it ROSAT} sources detected during pointed observations have
recently become available: the WGA catalogue (White et al. 1994)
and the ROSATSRC catalogue (Voges et al. 1994). These catalogues cover
about 10 per cent of the sky with a much higher sensitivity than that of the
RASS survey. Both catalogues include 50000 -- 60000 X-ray sources and
probably a few hundred BL Lacs. Identification of these sources has just
begun. Wolter et al. (in preparation) report the discovery of one BL Lac and
two BL Lac candidates. Giommi et al. (in preparation) also report the
identification of a WGA source with a BL Lacertae object.

\subsection{Gamma-ray catalogues}

The detection of a few BL Lacs at energies $\ga 100$ MeV by the Energetic
Gamma Ray Experiment Telescope (EGRET) on the {\it Compton Gamma Ray
Observatory}~({\it CGRO}; Gehrels, Chipman \& Kniffen 1993) has been reported
recently (Fichtel
et al. 1994). So far, only previously known BL Lac objects have been associated
with $\gamma$-ray sources, five at a high confidence level ($> 5 \sigma$),
four (including the uncertain BL Lac PKS 0521$-$365) at a lower (between
4$\sigma$) and 5$\sigma$) confidence level, and one (PKS 2155$-$304) for which
the
confidence level is not yet available (Vestrand, Stacy \& Sreekumar 1995).
(One of the five sources detected with a
high confidence level, S4 0954+658, has been detected only during phase 2 of
the EGRET observations [Mukherjee et al. 1995] and therefore is not included
in the EGRET phase 1 catalogue of Fichtel et al. 1994. This also applies to
PKS 2155$-$304). A cross-correlation of our BL Lac list
with the EGRET catalogue shows that some high-latitude marginal detections,
still unidentified, might be associated with BL Lacs (some of these are also
described in the notes to table 11B in Fichtel et al. 1994). These include: 3C
66A, which is 52 arcmin away from GRO J0222+42 (with which it had been
previously
identified), which has a 95 per cent error radius of 47 arcmin; MS1312.1$-$422,
only 17 arcmin away from GRO J1314$-$42, with an error radius of 71 arcmin.
This is
classified as a `possible' identification by Fichtel et al. (1994), probably
because its relatively low radio flux (18.5 mJy at 5 GHz) would make it the
extragalactic object with the largest $\gamma$-ray-to-radio flux ratio (all
the other
sources, in fact, have 5-GHz radio fluxes typically larger than 1 Jy). We
note, however, that within the 95 per cent error radius of GRO J1314$-$42
there are about 10 unclassified radio sources with $f_{\rm r} \ga 50$ mJy in
the Parkes-MIT-NRAO
(PMN) Southern survey (Wright et al. 1994), three of which have $f_{\rm r} >
100$ mJy. It is therefore likely that the counterpart of the $\gamma$-ray
detection is one of the brighter PMN sources. PKS 2032+107 is 89 arcmin away
from
GRO J2039+11, with an error radius of 66 arcmin, while PKS 2029+121 is somewhat
more distant, at 110 arcmin. Both objects have 5-GHz radio fluxes around 1 Jy.
PKS 2149+173 is 86 arcmin away from GRO J2157+18, with an error radius of 47
arcmin,
while the flat-spectrum radio quasar PKS 2201+171 is offset by 99 arcmin.
Again,
both objects have 5-Ghz radio fluxes around 1 Jy. The BL Lac candidate
1ES1745+504, with a 5-GHz radio flux as small as 1.3 mJy, was found to be
62 arcmin away from GRO J1742+49, with an error radius of 93 arcmin. However,
a much
more likely identification is that of the flat-spectrum radio quasar S4
1738+499, offset from the $\gamma$-ray source by 33 arcmin and with a 5-GHz
radio
flux of 0.6 Jy. Finally, a few BL Lacs are relatively near, although still
outside the 95 per cent error circle, to $\gamma$-ray sources mostly
identified with radio quasars. In some of these cases part of the $\gamma$-ray
emission might be due to the BL Lac.

\subsection{Miscellaneous objects}

Twenty-four objects have been classified as BL Lacs both by V\'eron-Cetty \&
V\'eron (1993a) and by Hewitt \& Burbidge (1993) but do not belong to any BL
Lac
sample. The reality of their classification has been recently confirmed by
V\'eron-Cetty \& V\'eron (1993b) for six of them: PKS 0047+023, PKS
0301$-$243, PKS 0808+019, PKS 1604+159, PKS 1717+177 and PKS 2254$-$204. We
have
no reason to suspect that the remaining objects would not satisfy the criteria
adopted in the definition of most BL Lac samples.

Five sources (PKS 0406+121, PKS 0422+004, PKS 0754+100, MC2 1307+12 and PKS
1413+135) have been reported as having radio fluxes at 5 GHz larger than 1 Jy
and
have $|b| > 10^{\circ}$ so they should in principle belong to the 1-Jy
catalogue (and therefore to the 1-Jy sample). The fact that they do not shows
that variability can have an effect even on radio samples.

\subsection{Uncertain and candidate BL Lacs}

Fifty objects are listed as uncertain or candidate BL Lacs. This list is quite
heterogeneous. It contains BL Lac candidates from various samples and found in
the literature. It also includes sources belonging to the compilation of
V\'eron-Cetty \& V\'eron (1993a) or of Hewitt \& Burbidge (1993) but not to
both.
Note that the $V$ magnitudes reported by V\'eron-Cetty \& V\'eron (1993a) are
actually $B$ magnitudes if no $B-V$ colour is given. In those cases, which
include also a few miscellaneous objects, we estimate the $V$ magnitude
assuming $B-V=0.6$, the mean value for BL Lacs.

\section{Statistical Properties}

The BL Lac catalogue assembles sources selected in different bands with
different flux limits. Therefore, detailed statistical analysis should be
restricted to appropriate subsamples and not to the whole collection of
objects, unless all the biases due to the different selection processes can
be fully taken into account. We can nevertheless have an overview of some
general properties like the distribution on the plane of the sky and the
redshift distribution.

\beginfigure{1}
\vskip 70mm
\caption{{\bf Figure 1.} The Aitoff projection in celestial (i.e.
equatorial) coordinates of all BL Lacs in the catalogue. Note the strong bias
in favour of northern declinations.}
\endfigure

Fig. 1 shows the Aitoff projection in equatorial coordinates of the BL Lacs in
the catalogue. A bias in favour of northern declinations is clearly present:
out of 233 objects, 178 (i.e. 76 per cent) have $\delta > 0^{\circ}$ and there
are no known BL Lacs with $\delta < -53^{\circ}$. This reflects the fact that
most major surveys have been done in the northern hemisphere, which has a
direct influence on radio samples and an indirect influence on serendipitous
X-ray samples.

\beginfigure{2}
\vskip 70mm
\caption{{\bf Figure 2.} Redshift distribution for all BL Lacs in
the catalogue. The hatched area indicates BL Lac candidates.}
\endfigure

Only 115 objects (or about 50 per cent of the catalogue), excluding lower
limits, have redshifts, which shows how difficult it is to extract this
important information from BL Lac spectra. Fig. 2 shows that the redshift
distribution for the catalogue peaks at $z \sim 0.1 - 0.2$. Note that,
apart from a BL Lac candidate at $z = 1.715$, the most distant confirmed
object reaches $z = 1.215$ and only four objects (plus two lower limits) have
$z > 1$. This is in marked contrast to the redshift distribution of quasars in
the Hewitt \& Burbidge (1993) catalogue (see their fig. 3) and is unlikely to
be due to selection effects inherent to the different sample. The redshift
distributions of BL Lacs and quasars, in fact, are also quite different in
complete samples: in the 1-Jy catalogue, for example (Stickel et al. 1994),
$z_{\rm max} \sim 3.8$ for quasars but only $\sim 1.2$ for BL Lacs. This is
probably related to the small (possibly absent or even negative) cosmological
evolution displayed by BL Lacs (see e.g. the discussion in Padovani \& Giommi
1995a).

The most `popular' BL Lac is Mrk 421 alias S4 1101+364 alias PG 1101+385
alias 1H 1104+382 alias 1ES1101+384 alias GRO J1106+38, as it belongs to five
samples: S4, PG, {\it HEAO}-1, Slew, and {\it GRO}. It is closely followed by
three other BL Lacs which belong to four samples: PG 1218+304 alias
EXO1218.8+3027 alias 1H 1219+301 alias 1ES1218+304 (PG, {\it EXOSAT}, {\it
HEAO}-1,
and Slew surveys); Mrk 501 alias S4 1652+398 alias 1H 1651+398 alias
1ES1652+398 (1-Jy, S4, {\it HEAO}-1, and Slew surveys); and 3C 371 alias S4
1807+698 alias 1H 1803+696 alias 1ES1807+698 (1-Jy, S4, {\it HEAO}-1, and Slew
surveys).

As anticipated in the Introduction, a first use of this catalogue has been the
extraction of all BL Lacs from the WGA catalogue to study their X-ray
properties. Those results will be presented elsewhere (Padovani \& Giommi
1995b).

ASCII and \TeX~versions of the catalogue can be obtained on NCSA MOSAIC at
the following URL: {\tt
http\-://itovf2.roma2.infn.it/pa\-do\-va\-ni\-/ca\-ta\-logue\-.html}.
We welcome comments, suggestions, corrections and additions, all of which
should
be addressed to the first author (electronic mail:
padovani\-@roma2.\-infn.\-it).

\section*{Acknowledgments}

We thank Richard Green for confirming the identification of Mrk 421 with
PG 1101+385, Eric Perlman for providing us with the latest Slew survey BL Lac
list in advance of publication, Joe Pesce for useful comments, and Manfred
Stickel for information about the S5 BL Lacs. This research has made use of
the BROWSE program developed by the
ESA/{\it EXOSAT} Observatory and by NASA/HEASARC, and of the on-line services
provided by the European Space Information System (ESIS), and by the NASA/IPAC
Extragalactic Database (NED), which is operated by the Jet Propulsion
Laboratory, California Institute of Technology, under contract with the
National Aeronautics and Space Administration.

\bye